\pdfoutput=1
\documentclass[superscriptaddress, 12pt, aps, prl,  preprint, longbibliography]{revtex4-1}
\usepackage{graphicx}
\usepackage{color, soul}
\usepackage{epstopdf}
\usepackage{amsmath, amssymb}

\begin{document}

\title{Formation of abiogenic hydrocarbons in supercritical fluids under Earth's upper mantle conditions}
\author{Nore Stolte}
\altaffiliation[Present address: ]{Lehrstuhl f{\"u}r Theoretische Chemie, Ruhr-Universit{\"a}t Bochum, 44780 Bochum, Germany}
\affiliation{Department of Physics, Hong Kong University of Science and Technology, Hong Kong, China}
\author{Tao Li}
\email{bruceli@ust.hk}
\affiliation{Department of Physics, Hong Kong University of Science and Technology, Hong Kong, China}
\author{Ding Pan}
\email{dingpan@ust.hk}
\affiliation{Department of Physics, Hong Kong University of Science and Technology, Hong Kong, China}
\affiliation{Department of Chemistry, Hong Kong University of Science and Technology, Hong Kong, China}

\date{\today}
\begin{abstract}
The formation of hydrocarbons in Earth’s interior has traditionally been considered to have biogenic origins; 
however, growing evidence suggests that some hydrocarbons may instead originate abiotically in the deep carbon cycle. 
It is widely expected that the Fisher-Tropsch-type (FTT) process, which typically refers to the conversion of inorganic carbon to organic matter in 
geological settings,
may also happen in Earth's interior, but the absence of industrial catalysts and aqueous conditions in deep environments suggest that the FTT process can be very different from that in the chemical industry.
Here, we performed extensive \textit{ab initio} molecular dynamics (AIMD) simulations ($>$ 2.4 ns) to investigate the FTT synthesis 
in dry mixture and in aqueous solutions at 10-13 GPa and 1000-1400 K.
We found that large hydrocarbon-related species containing C, O, and H
($>$C$_2$)
are abiotically synthesized via the polymerization of CO without any catalyst. 
Supercritical water, commonly found in 
the
deep Earth, does not prevent organic molecule formation but restricts product size and carbon
reduction.
Our studies reveal a previously unrecognized abiogenic route for hydrocarbon synthesis in mantle geofluids. These carbon-containing fluids could potentially migrate from depth to shallower crustal reservoirs, thereby influencing Earth's surface carbon budget.

\end{abstract}
\maketitle

\section{Introduction}
Recently, hydrocarbons potentially originating through abiotic processes have been discovered in fluids from many different geological environments \cite{Sephton2013On, McCollom2013Laboratory, vitale2017massive, Tao2018Formation, Sverjensky2020Changing, vitale2020subduction, pena2021situ}.
The formation of these hydrocarbons from abiogenic precursors in Earth's interior plays a role in the deep carbon cycle, which substantially influences the carbon budget near Earth’s
surface and has significant implications for global climate change over geological time \cite{Kelemen2015Reevaluating, Kutcherov2020Hydrocarbon, Hazen2013Deep}.
While the abiogenic formation of petroleum was first proposed by Mendeleev in the 19th century \cite{Mendeleev1877}, serious scientific investigation into hydrocarbon generation from subducting carbon and water—a part of the deep carbon cycle—has only emerged in recent decades \cite{Gold1980Deep, Kolesnikov2009Methane, kutcherov2010synthesis, Sephton2013On, McCollom2013Laboratory, kolesnikov2017chemistry, Kutcherov2020Hydrocarbon}. The chemical process by which hydrocarbons may form under the extreme conditions of Earth's upper mantle is still largely unknown.

While abiotic CH$_4$ is considered as a major carbon species in geofluids where strongly reducing conditions prevail \cite{vitale2017massive, Tao2018Formation, vitale2020subduction,pena2021situ}, recent geological evidence and experimental studies suggest that higher hydrocarbons  
($>$C$_2$)
may also exist in deep Earth \cite{Sephton2013On, Sverjensky2020Changing}. 
In chemical engineering, the Fisher-Tropsch (FT) process is an important synthesis method to produce heavy hydrocarbons  \cite{Brady1981Mechanism}. It includes a series of catalytic reactions converting CO and H$_2$ into heavier hydrocarbons:
\begin{equation}
(2n + 1)\,\mathrm{H}_2 + n\,\mathrm{CO} \rightarrow \mathrm{C}_n\mathrm{H}_{2n+2} + n\,\mathrm{H}_2\mathrm{O}.
\end{equation}
Many geochemists anticipate that such reactions may also occur in deep Earth; however, industrial catalysts such as transition metal alloys typically do not exist in natural environments. In the relevant studies, the Fischer-Tropsch-type (FTT) synthesis typically refers to the conversion of inorganic carbon to organic matter in a broader context \cite{McCollom2013Laboratory}.
Another key difference between industrial and geochemical FTT synthesis is the presence of water. Despite the production of a small amount of water, industrial FTT reactions typically occur in dry gas mixtures without any liquid water, whereas
in subsurface geologic environments, the FTT process occurs in the medium of liquid or supercritical water \cite{McCollom2013Laboratory}. 
Water molecules are known to inhibit or promote the catalytic conversion of small carbon-containing molecules in the FT synthesis \cite{Jiang2024Role};
however, the influence of the aqueous environment on the uncatalyzed FTT process remains poorly understood.

In Earth's upper mantle, 
pressure (P) and temperature (T) reach approximately 13 GPa and 1700 K, respectively \cite{Thompson1992Water}.
Many recent geochemical studies provide compelling evidence for a free aqueous fluid phase in the upper mantle \cite{Manning2013Chemistry, Tschauner2018Ice-VII, Sverjensky2020Changing,bali2013water}.
Geological fluids in the upper mantle have traditionally been simulated as simple mixtures of small volatile molecules, e.g., H$_2$O, CO$_2$, CO, CH$_4$, H$_2$ \cite{Zhang2009Model};
however, recent experimental and theoretical studies have revealed the significant roles played by 
chemical reactions 
at supercritical conditions in Earth’s lithosphere
\cite{Pan2013Dielectric, 
Sverjensky2014Important,
Pan2016Fate, Abramson2017Water-carbon, kolesnikov2017chemistry, Stolte2019Large, 
Sverjensky2020Changing, 
dettori2020carbon, stolte2021water, stolte2022nanoconfinement, li2024synthesis, fei2025abiotic}.
Experimental investigation of reactions under such extreme P-T conditions remains highly challenging \cite{Kolesnikov2009Methane, Lobanov2013Carbon, McCollom2013Laboratory}.
First-principles atomistic simulations, which require neither experimental input nor empirical parameters, have been successfully applied to study chemistry under extreme conditions, providing crucial molecular-level insights \cite{Gygi2005Abinitio}.
Spanu et al. applied \textit{ab initio} molecular dynamics (AIMD) simulations to show that larger hydrocarbons are thermodynamically favored in Earth's deep interior \cite{Spanu2011Stability}. However, the reaction pathways for methane conversion to higher hydrocarbons still require clarification.
Kuang and Tse applied AIMD to study reactions between H$_2$ and CaCO$_3$ under extreme P-T conditions corresponding to Earth's lower mantle and core-mantle boundary, observing the formation of tetrahedral C$_4$ moieties and water~\cite{kuang2022high}.
Additional studies have explored methane reactions and stability at megabar pressures and temperatures of several thousand kelvins---conditions more extreme than Earth's mantle and typical of ice giant planets like Neptune and Uranus \cite{ancilotto1997dissociation, lee2011mixtures, li2011quantum}.
Nevertheless, our understanding of abiogenic hydrocarbon reactions under upper mantle conditions remains very limited.

In this work, we studied the abiogenic synthesis of hydrocarbons at 10--13 GPa and 1000--1400 K,
P-T conditions found in Earth's upper mantle.
We conducted extensive 
AIMD 
simulations, with a cumulative duration exceeding 2.4 ns, to investigate the uncatalyzed reactions of the FTT reactants CO + H$_2$ in a dry mixture and in aqueous solutions.
We found that large hydrocarbon-related species containing C, O, and H
are
abiotically synthesized via the polymerization of CO both in dry and aqueous mixtures, and that reactions of polymers with H$_2$ lead to reduction of carbon.
Higher pressure 
leads to
formation of
more organic species and larger products. 
Water does not prevent organic molecule formation but restricts product size and carbon reduction.
Finally, we discussed reaction mechanisms of abiogenic organic synthesis and implications to Earth's deep carbon cycle. 

\section{Results}
We first carried out AIMD simulations of a 1:1 supercritical mixture of CO and H$_2$ at 13~GPa and 1400~K. 
Then, we added water to the mixture, and performed AIMD simulations of 1:1:1 supercritical mixtures of CO, H$_2$, and H$_2$O at 10--13 GPa and 1000--1400 K (Table SI in the supporting information).
At each condition, we performed six independent 50 ps long (after equilibration) NVT simulations, giving a total of 300 ps of simulation at equilibrium for each condition.
We analyzed the whole trajectories to uncover reaction mechanisms.

\begin{figure}
\centering
\includegraphics[width=0.5\textwidth]{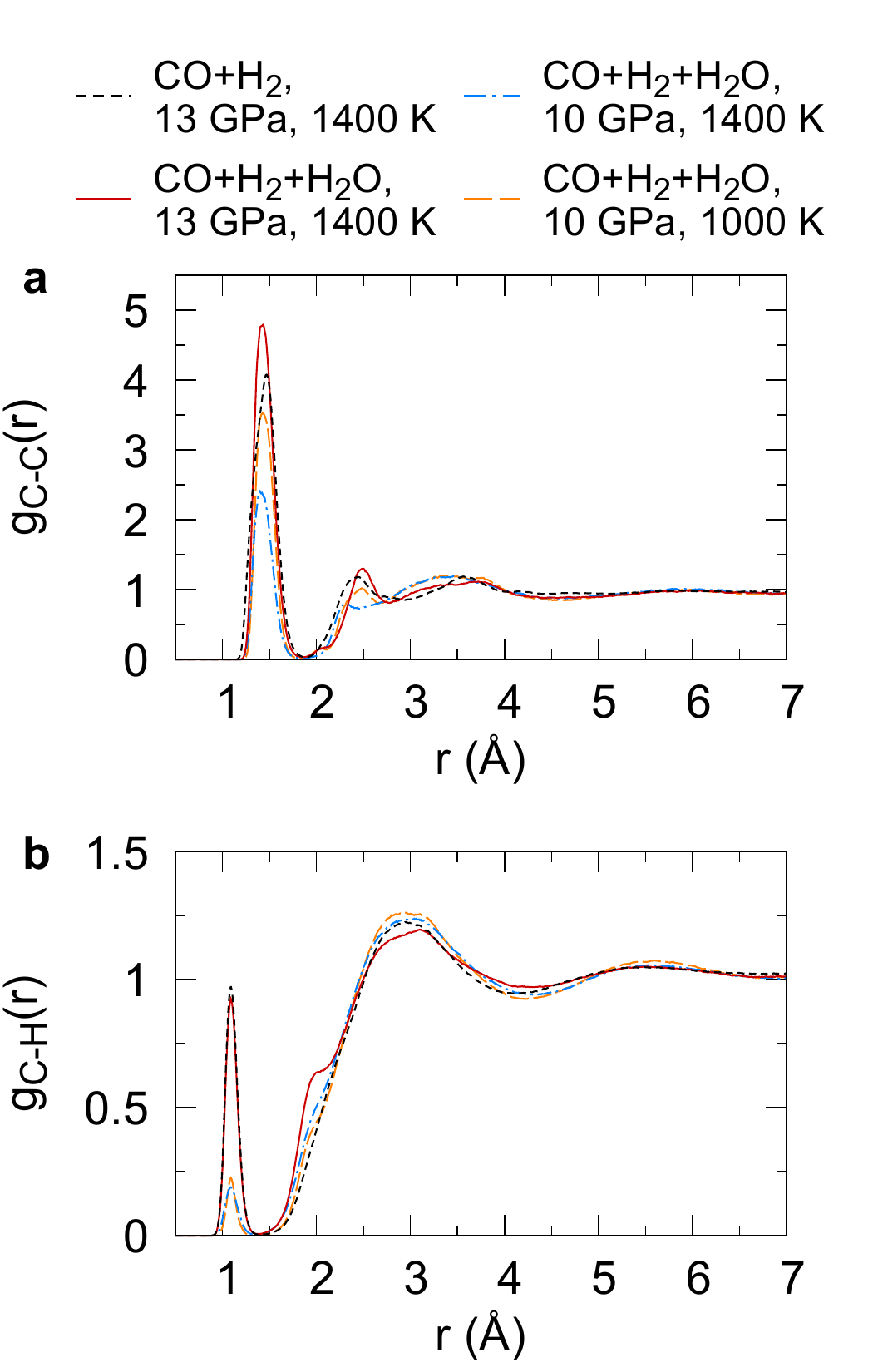}
\caption{Carbon-carbon (\textbf{a}) and carbon-hydrogen (\textbf{b}) radial distribution functions for mixtures of CO, H$_2$ and H$_2$O.}
\label{rdf-NVT}
\end{figure}

In all systems studied, the C-C and C-H radial distribution functions (RDFs) show a pronounced peak at covalent-bonding distances (Fig. \ref{rdf-NVT}), even though initially, there were only CO, H$_2$ and H$_2$O molecules in the supercritical mixtures.
The C-C RDFs have a maximum at 1.40--1.47 {\AA}, indicating that there is likely a mixture of C-C single bonds (bond length $\sim$1.54 {\AA}) and C=C double bonds (bond length $\sim$1.34 {\AA}) present in the fluids.
The C-H RDFs have the first maximum at 1.09 {\AA}. The C-H bond length in alkanes or alkenes is $\sim$1.09 {\AA}, so we can conclude that C-H covalent bonds form in the supercritical mixtures.
Also, the C-C RDFs show enhanced structuring beyond the first peak, in particular at 13 GPa and 1400 K, indicating that extended molecules are present.

\begin{figure}
\centering
\includegraphics[width=0.5\textwidth]{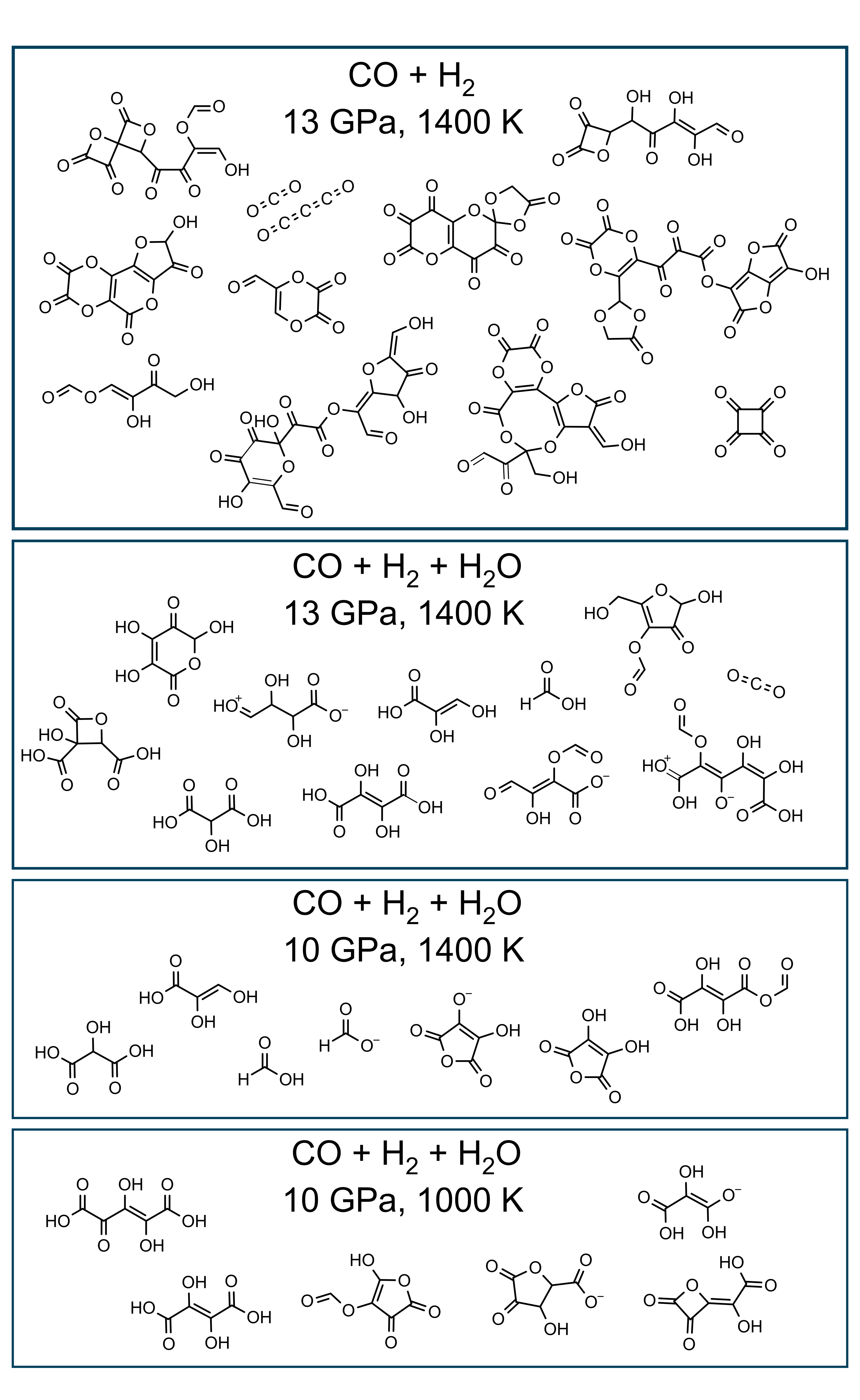}
\caption{
Reaction products in mixtures of CO, H$_2$ and H$_2$O at different P-T conditions.  
In addition to these products, a substantial amount of CO, H$_2$, and H$_2$O remained.}
\label{structures}
\end{figure}

\begin{figure}
\centering
\includegraphics[width=0.5\textwidth]{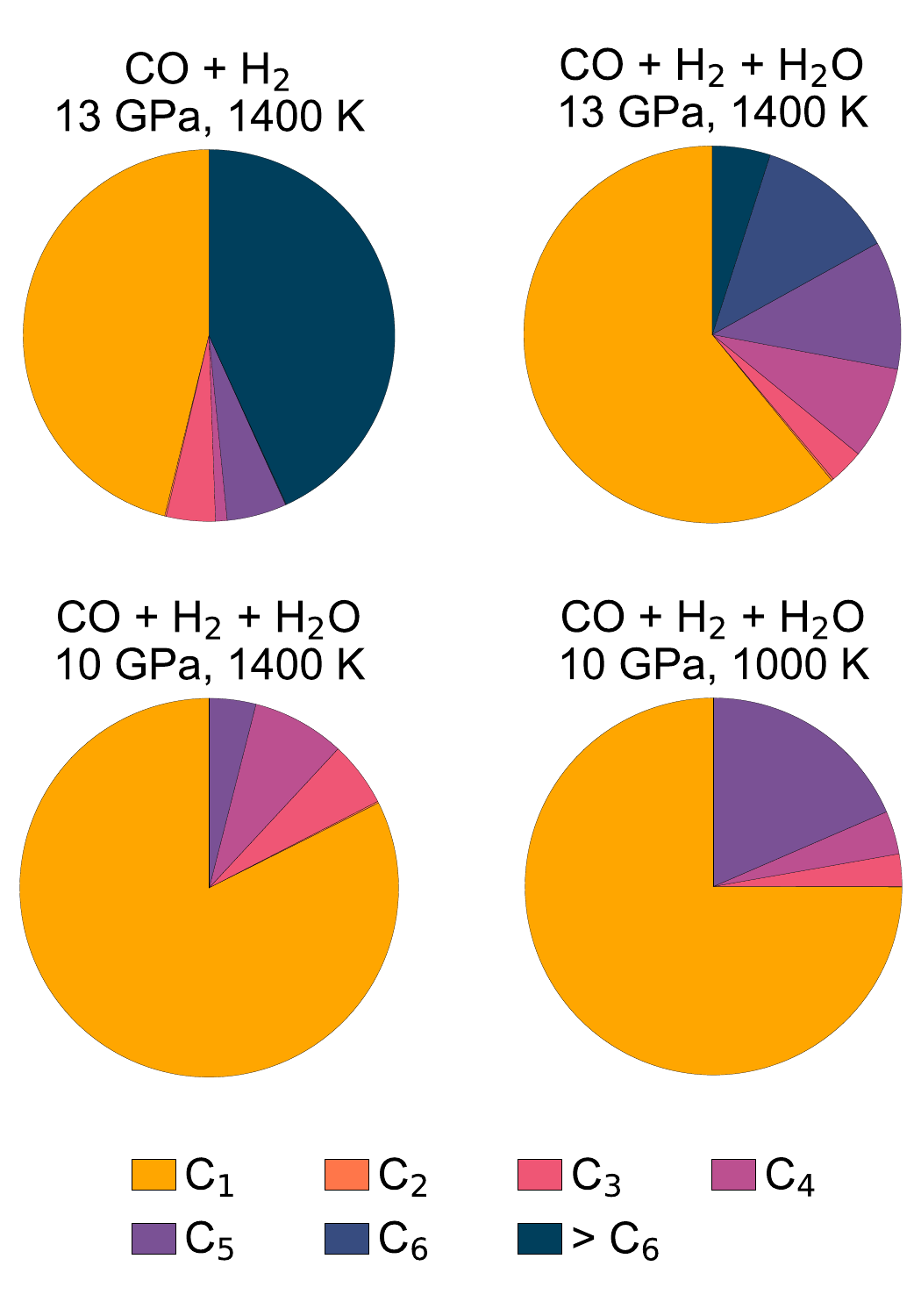}
\caption{Fraction of total carbon in C$_n$ molecules produced in mixtures of CO, H$_2$ and H$_2$O. $n$ is the number of carbon atoms in the molecule.}
\label{carbon_numbers_nvt}
\end{figure}

The overview of reaction products in Fig. \ref{structures} indeed
shows
that larger products form at larger pressure, and that products are larger in the absence of water.
At equilibrium in the dry CO + H$_2$ mixture at 13 GPa and 1400 K, more than half the CO molecules
had
reacted to form molecules containing three or more C atoms, and 43 \% of all carbon was contained in molecules with 7 or more C atoms (Fig. \ref{carbon_numbers_nvt}).
Typically, product molecules of the reactions between CO and H$_2$ at these conditions contain carbonyl bonds, ether bonds, hydroxyl bonds, C-H bonds, and rings incorporating both oxygen and carbon atoms.
Some minor C$_1$ products formed as well, namely carbon dioxide and formaldehyde, but CO molecules still make up the majority of C$_1$ molecules.

After adding water to the supercritical mixture of CO and H$_2$ at 13 GPa and 1400 K, polymerization of carbon still took place, but at equilibrium the carbon-containing molecules are smaller than in the dry mixture.
C$_1$ molecules make up 61 \% of the total C content at 13 GPa and 1400 K in the CO + H$_2$ + H$_2$O mixture (Fig. \ref{carbon_numbers_nvt}).
Carbon monoxide remains the most important C$_1$ molecule, with some CO$_2$, HCOOH and HCOO$^-$ species forming as well.
About 5 \% of the carbon is contained in molecules with more than 6 C atoms, and C$_4$, C$_5$ and C$_6$ molecules are more abundant in the aqueous mixture than in the dry one.
This indicates that water does not inhibit polymerization of CO at extreme conditions, but it does affect the size of the reaction products.

When we decreased the pressure to 10 GPa at 1400 K for the CO + H$_2$ + H$_2$O mixture, carbon-containing molecules decreased in size, and no molecules larger than C$_5$ were formed (Fig. \ref{carbon_numbers_nvt}).
Additionally, at 10 GPa the first peak in the C-H RDF is significantly smaller than at 13 GPa, i.e., decreasing pressure leads to formation of fewer C-H bonds (Fig. S5). 
The effect of pressure can be understood through the reaction stoichiometry. In polymerization reactions, several small reactant molecules react to form one large product molecule, so the effective volume of the reaction product is smaller than that of the reactants.
The Gibb's free energy change of reaction is given by $\Delta U + \mathrm{P}\Delta V - \mathrm{T} \Delta S$, where $U$ is the internal energy, $V$ is the volume, and $S$ is the entropy.
For polymerization reactions, P$\Delta V$ is negative, and it is more negative at larger pressure.
Therefore, product molecules are smaller and C$_1$ molecules are more abundant at 10 GPa than at 13 GPa at the same temperature.

Finally, we decreased the temperature from 1400 K to 1000 K at 10 GPa for CO + H$_2$ + H$_2$O.
With decreasing temperature, the first peak in the C-C RDF increased in height, meaning that more C-C bonds form at lower temperatures.
At 1000 K, C$_1$ molecules make up a smaller fraction of the total carbon content than at 1400 K, and more C$_5$ molecules exist at lower temperature (Fig. \ref{carbon_numbers_nvt}).
Overall, larger molecules form at 1000 K than at 1400 K at the same pressure because of entropic effects.
Less entropy is associated with large molecules than with small ones, so for polymerization reactions, T$\Delta S$ is negative, i.e., there is a free energy penalty associated with formation of large molecules.
The free energy penalty increases with temperature, so larger molecules are more stable at lower temperature.

In the solutions studied here, reactions that led to polymerization of carbon were initiated when a C-C bond formed between CO molecules.
A chain of CO molecules bonded through the carbon atoms is thermodynamically unstable relative to separated CO molecules \cite{Schroder1998Ethylenedione, Xin2019Global, Mato2020Stability}, and in our simulations such chains tended to react further within a few picoseconds.
In some cases, H$_2$ dissociated to form C-H bonds with carbon atoms and C-OH bonds with oxygen atoms in the CO polymers, which stabilized molecules as bonds became saturated.
If a polymer was not stabilized by termination with H$_2$, it tended to react with further CO molecules, which increased the size of the molecule, as shown in Fig. S6.

Alternatively, molecules reached stable configurations by forming a ring, as can be seen in Fig. \ref{structures}.
At equilibrium, these rings have 4 to 8 vertices, and typically incorporate at least one oxygen atom, such as substituted oxetane, (di)oxolane, and (di)oxane rings.
We did not find any three-membered rings, which have quite a strained geometry.
The most extreme case of stabilization of a polymer through the formation of a ring that we observed in our simulations was the formation of neutral cyclobutane-1,2,3,4-tetrone (C$_4$O$_4$) from 4 CO molecules in the CO + H$_2$ mixture at 13 GPa and 1400 K (Fig. S7 in the supporting information).
In the gas phase, C$_4$O$_4$ is thermodynamically unstable relative to 4 CO molecules \cite{Gleiter1995Stability}, though it has been studied using negative ion photoelectron spectroscopy \cite{Guo2012Probing}.
In supercritical mixture of CO and H$_2$, C$_4$O$_4$ in one case had a lifetime exceeding 20 ps, so it is possible that the high pressure leads to stabilization of the triplet state molecule.

\begin{figure}
\centering
\includegraphics[width=1.0\textwidth]{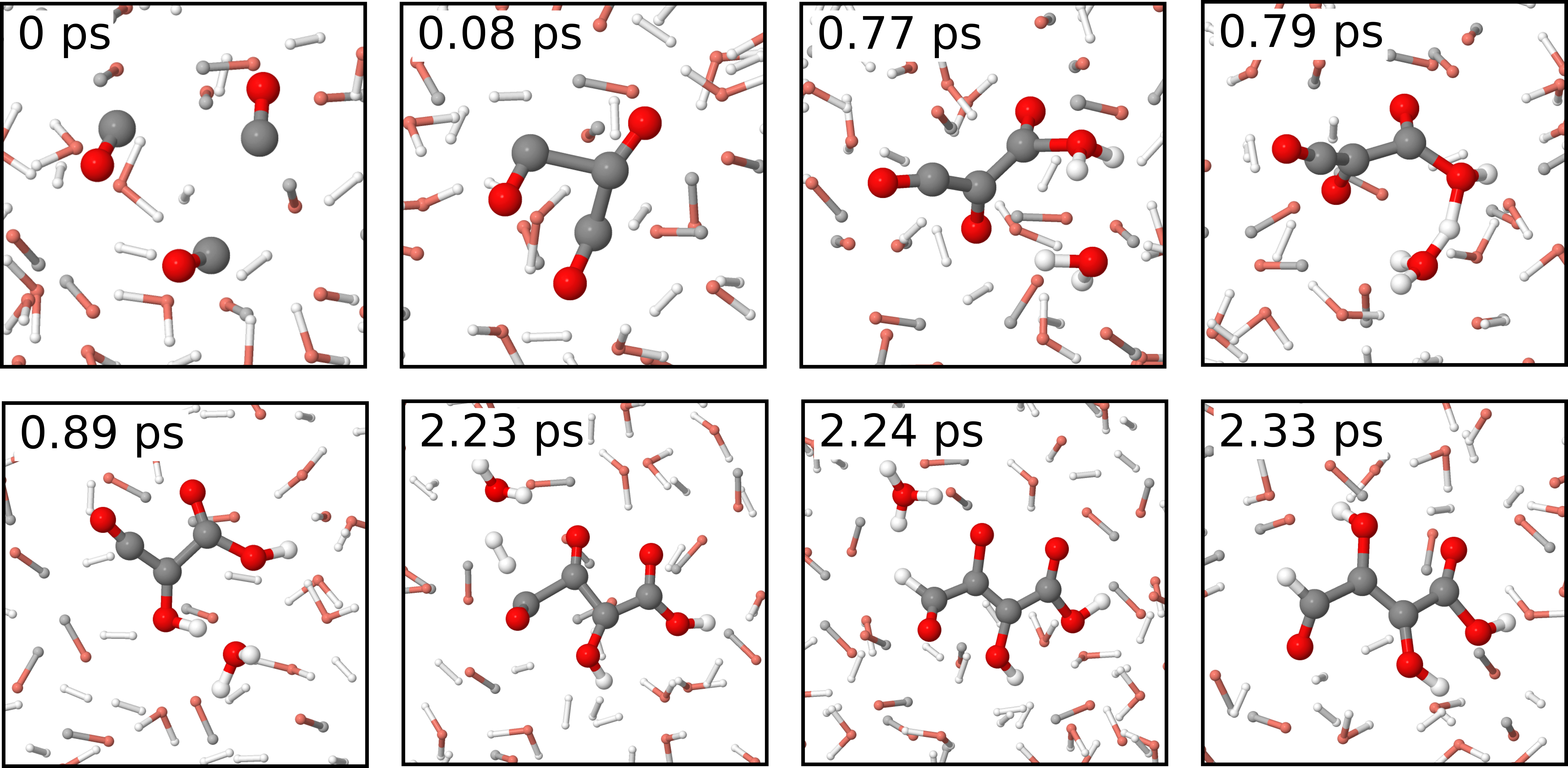}
\caption{
Formation of 2,3-hydroxy-4-oxobut-2-enoic acid from CO, H$_2$ and H$_2$O at 13 GPa and 1400 K.}
\label{mechanism}
\end{figure}

When water was present, the molecules could be stabilized in a further way as well. In addition to reactions with H$_2$ and formation of rings, (CO)$_n$ polymers were stabilized through reactions with H$_2$O.
H$_2$O can donate H$^+$ or OH$^-$, or both, to the polymer, forming C-H and C-OH bonds. 
Unstable CO chains readily react with further CO molecules, thereby increasing the size of the molecule, but molecules that are saturated by C-H or C-OH bonds are not so reactive.
Therefore, water affects the carbon polymerization reactions in CO + H$_2$ + H$_2$O solutions by saturating bonds in the unstable CO chains, which prevents further CO molecules from bonding to the polymer, resulting in smaller reaction products.
The stable end groups on the carbon polymers formed by reactions with water or H$_2$ also prevent rings from forming in the molecule, so cyclic compounds are less abundant in the presence of water (Fig. \ref{structures}).
A CO polymerization reaction mechanism with water is illustrated in Fig.~\ref{mechanism}. Initially, three CO molecules bond through their carbon atoms. Water dissociates to form a new hydroxyl group at a terminal carbon atom, while donating the second hydrogen atom to the central CO unit, forming another hydroxyl group. The carbon atom at the other end reacts with CO, which increases the size of the molecule to C$_4$. At this end, a H$_2$ atom dissociates, and one of its atoms forms a C-H bond. The second atom is donated to an oxygen atom of the same molecule with the participation of a water molecule. This leaves 2,3-hydroxy-4-oxobut-2-enoic acid. A CO polymerization reaction in the dry CO + H$_2$ mixture is illustrated in Fig. S7 in the supporting information.

One final way in which CO polymers reached a stable configuration was through the breaking of the carbon monoxide C-O bonds.
Although this happened rarely in our simulations, we observed a few instances where three CO molecules bonded through the carbon atoms, and the central C-O bond broke to leave a stable linear carbon suboxide molecule, C$_3$O$_2$ (Fig. S8 in the supporting information).
In the industrial FT synthesis, the breaking of the C-O bond of carbon monoxide at the catalyst surface is a central step in the reaction pathway, so the fact that C-O bonds broke so rarely in reactions studied here illustrates that the reaction mechanisms that we observed are distinct from those involved in the industrial FT synthesis.

Another remarkable observation is that no C$_2$ molecules formed during our NVT simulations (Fig. \ref{carbon_numbers_nvt}), both with and without water.
The dimer of CO (ethylene dione) is energetically unstable with respect to dissociation to two CO molecules even though its ground state is a triplet state, as the singlet-triplet crossing point occurs close to the triplet energy minimum with respect to the bend angle \cite{Schroder1998Ethylenedione, Mato2020Stability}.
As a result, any CO molecules that do temporarily form a covalent OC-CO bond, as defined by the C-C distance, rapidly dissociate again, before the molecule can be stabilized in further reactions.

\begin{figure}
\centering
\includegraphics[width=0.5\textwidth]{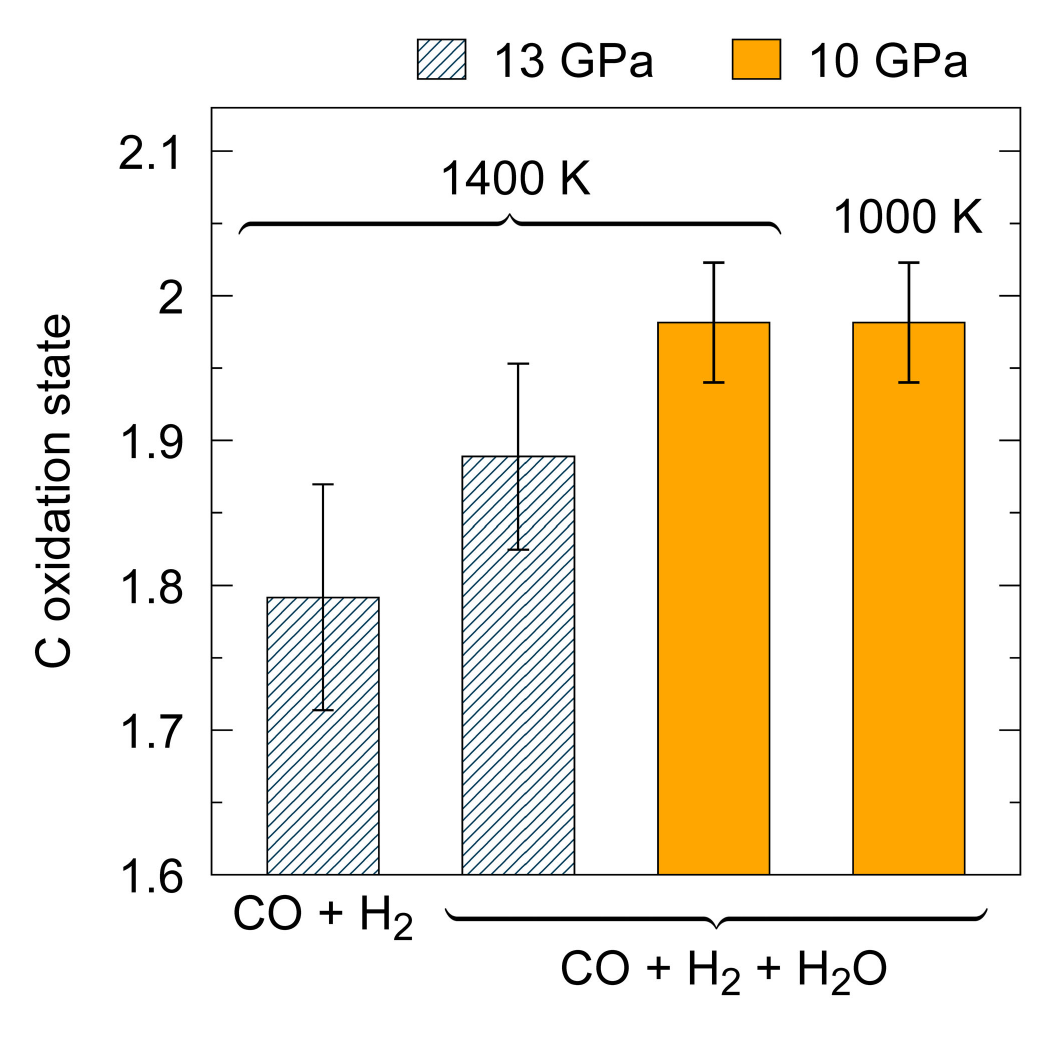}
\caption{The mean oxidation state of carbon atoms in the different mixtures of CO, H$_2$ and H$_2$O. The hatched bars show data at 13 GPa, and the solid bars show data at 10 GPa. Temperature is indicated above the bars, and composition is indicated below the bars. The local charge on C atoms and their bonded neighbors was determined using the maximally localized Wannier function centers to localize electrons \cite{Marzari2012Maximally}, so that the oxidation state of carbon could be deduced. The error bars show the standard deviation of the mean oxidation state at each time step in simulations.}
\label{oxidation}
\end{figure}

The oxidation state of carbon in CO is +2, but formation of new C-O and C-H bonds alters the carbon oxidation state.
In the dry CO + H$_2$ mixture at 13 GPa and 1400 K, the 
average
oxidation state of carbon
in the system
was reduced to +1.79 $\pm$ 0.08 at 13 GPa, 1400 K (Fig. \ref{oxidation}).
The main cause of this reduction is the formation of C-H bonds, rather than the breaking of C-O bonds. As discussed, we rarely observed breaking of the C-O bond, and when a C-O bond was broken, it always led to the formation of CO$_2$, where carbon has an oxidation state of +4.
In the presence of water, the average oxidation state of carbon is +1.89 $\pm$ 0.06 at 13 GPa and 1400 K.
In aqueous solutions, not only did C-O bonds break, but new C-O bonds formed as well, because H$_2$O reacted with CO molecules, forming CO$_2$, HCOOH, and oxygen-rich polymers.
Therefore, carbon is more oxidized when water is present in the reaction mixture, though carbon is still reduced relative to its oxidation state in CO.
The average oxidation state of carbon atoms is only slightly below +2 for the CO + H$_2$ + H$_2$O mixture at 10 GPa and 1400 K.
This slight reduction of carbon is due to the formation of C-H bonds, which occurs less readily at 10 GPa than at 13 GPa (Fig. S5 in the supporting information).
The oxidation state of carbon does not change when going from 1400 K to 1000 K at 10 GPa, because the number of new C-O and C-H bonds that forms at both conditions is nearly equal (Fig. S5 in the supporting information).
Overall, we find that at the conditions studied, temperature does not substantially affect the oxidation state of carbon, but increasing pressure leads to the reduction of carbon, primarily as a result of the formation of C-H bonds.
In the presence of water, the formation of new C-O bonds results in more oxidized carbon products than in the dry CO + H$_2$ mixture at the same P-T conditions.

\section{Discussion}
Abiogenic synthesis of hydrocarbons from supercritical C-O-H fluids could take place under Earth's upper mantle conditions, even without any catalyst. 
The carbon polymerization reaction mechanism in our study is distinctly different from the FTT synthesis 
that occurs
in hydrothermal environments \cite{mccollom1999lipid,McCollom2006Carbon,McCollom2013Laboratory}, because the extreme pressure plays an important role.
Higher pressures lead to increased stabilization of large molecules for addition reactions, and more organic species and larger size of hydrocarbons (\textgreater{C$_6$})
are formed at 13 GPa. 
During the synthesis process, carbon in CO does not need to be reduced before a polymer forms, and in fact, H$_2$ molecules dissociate and react with the polymer of CO molecules only \emph{after} the reaction is initialized through formation of C-C bonds. 
More importantly for Earth's upper mantle environment, we find that water does not inhibit C–C bond formation, as evidenced by the generation of molecules containing up to five C–C bonds in our simulations.
However, carbon remain more oxidized in aqueous solutions than in dry mixtures of CO and H$_2$ because reactions with water lead to formation of new C-O bonds, and these hydrocarbon-related products have larger oxygen content. Higher pressure is conducive to the reduction of carbon due to the formation of C-H bonds. 
These organic species are notably characterized by a significant presence of oxygen atoms.

Our findings reveal a new pathway for the abiotic synthesis of higher hydrocarbons in upper mantle geofluids.
We showed that carbon in the form of CO can polymerize in water-rich regions of the mantle, such as subduction zones. These zones are critical to the deep carbon cycle, as downgoing tectonic plates deliver vast amounts of carbon from Earth's surface to its interior \cite{Kelemen2015Reevaluating, Plank2019Subducting}.
The transport of hydrocarbons in subduction zones may greatly affect the geological environment in deep Earth \cite{Sverjensky2014Important,giuntoli2024methane, dobe2025tracking}. Furthermore, the subsequent rise of hydrocarbons can significantly influence the carbon budget near Earth's surface.
Thus, the abiotic formation of hydrocarbons we found represents a potentially important component of the deep carbon cycle.

\section{Conclusion}
In summary, we studied the FTT synthesis under aqueous conditions without catalysts. 
our extensive AIMD simulations (\textgreater 2.4 ns) show that hydrocarbons and related organic species can be abiotically synthesized under upper mantle conditions (10–13 GPa, 1000–1400 K) through CO polymerization, even in the absence of catalysts. Different from the FT process in industrial environments, extreme pressure promotes C–C bond formation 
and stabilizes larger hydrocarbons (\textgreater{C$_6$}). 
Supercritical water is common in Earth's deep interior. While it does not inhibit organic synthesis, it limits product size and carbon reduction compared to dry CO-H$_2$ mixtures.
Our study highlights a previously unrecognized route for abiotic hydrocarbon synthesis in Earth’s deep interior, with important implications for the deep carbon cycle.

\section*{Methods}
\subsection{Ab initio molecular dynamics (AIMD) simulations}
We carried out Born-Oppenheimer AIMD simulations using the Qbox code \cite{Gygi2008Architecture}. 
Here, we have used the PBE exchange-correlation functional \cite{Perdew1996Generalized}.
Previously, we found that the discrepancies between results obtained with the PBE and the hybrid PBE0 functional~\cite{Adamo1999Toward} are significantly smaller at extreme P-T conditions than at ambient conditions~\cite{Pan2013Dielectric, Pan2014Refractive, Pan2016Fate}. 
For example, the PBE and PBE0 results for speciation of CO$_2$(aq) at upper mantle conditions are qualitatively equivalent~\cite{Pan2016Fate}.
Because the PBE and PBE0 functionals do not adequately describe van der Waals (vdW) interactions, we additionally tested vdW corrections using Grimme's D3 method \cite{grimme2010consistent} with Becke-Johnson damping implemented in the RPBE-D3 functional \cite{hammer1999improved}. We found these corrections produced only minor variations ($<$6\%) in carbon species mole fractions \cite{stolte2022nanoconfinement}. This confirms that dispersion interactions play an insignificant role in covalent bond transformations and consequently have negligible effects on the observed chemical speciation.
Therefore, we expect that our results for the carbon speciation at extreme P-T conditions are robust with respect to the choice of functional.

We employed a plane-wave basis with ONCV SG15 pseudopotentials~\cite{Hamann2013Optimized, Schlipf2015Optimization}.
The time step was 0.24 fs.
We used deuterium atoms instead of hydrogen atoms in simulations, which allowed for the use of the larger time step.
For ease of interpretation of our results, we still refer to these atoms as hydrogen atoms in the text.
Temperature in simulations was maintained using the Bussi-Donadio-Parrinello thermostat~\cite{Bussi2007Canonical}, with a relaxation time of 24.2~fs. 
We performed constant pressure (NPT) simulations until the unit cell volume converged. In these variable-cell AIMD simulations, the plane-wave energy cutoff was set to 85~Ry, with the reference cell energy cutoff set to 70~Ry \cite{focher1994structural}.
After the unit cell volume converged at the desired P-T conditions, a new unit cell was constructed to start constant volume (NVT) simulations, where the plane-wave energy cutoff was 65~Ry. The cutoff was increased to 85~Ry to verify the pressure.
The pressure in NVT simulations was computed before reactions occurred in the simulation, i.e., for the mixture of reactants.
In the supporting information, we show the potential energy as a function of time (Fig. S1--S4).
We considered the simulations equilibrated when there were no longer large fluctuations in the total energy.
The equilibration time was different for each run, depending on the initial configuration of the system.
Table SI in the supporting information summarizes the simulation set-ups.

\subsection*{Structure determination}
The atoms in the first peak in the C-C and C-O RDFs form covalent bonds with the reference C atom.
Therefore, the first minimum in the RDF, obtained from the production runs of NVT simulations, is used as the cutoff distance for C-C and C-O bonds at each P-T condition.
Table SII in the supporting information lists all the cutoff distances.
Each hydrogen atom is considered bonded to its nearest neighbor, unless the nearest neighbor is another hydrogen atom bonded to a third atom.
In such cases, the hydrogen instead bonds to its second-nearest neighbor.
This allows us to determine the atoms and connectivity in each molecule.

\subsection{Oxidation state}
To calculate the oxidation state of carbon atoms, we computed their local charge, defined as the sum of the nuclear charge (+4$e$ for carbon, considering only valence electrons) and the electronic charge associated with each carbon atom, based on the assumption that all oxygen atoms have an oxidation state of -2, and all hydrogen atoms have an oxidation state of +1.
The local electronic charge, $q_{\rm local}$, was obtained by tracking the maximally localized Wannier function (MLWF) centers \cite{Marzari2012Maximally},
which each correspond to an electron pair with charge -2$e$.
For each NVT trajectory, we performed an extra 5 ps of simulation in which we computed the 
MLWF
centers every 10 time steps. 
The MLWF centers, denoted M, 
were considered to be localized on their nearest-neighbor atom.
MLWF centers were assigned to a carbon-carbon bond between two carbon atoms C$^1$ and C$^2$ if both the M-C$^1$ and M-C$^2$ distances were smaller than the C$^1$-C$^2$ distance, and the C$^1$-M-C$^2$ angle was larger than 111$^{\circ}$.
The choice of angle was the smallest possible angle that avoided ambiguity in most of the assignments of MLWF centers to C-C bonds. 
This determined C-C single, double, and triple bonds.
The only exception was cyclobutane-1,2,3,4-tetrone (C$_4$O$_4$), where MLWF centers sometimes drifted towards the center of the four-membered ring leading to inconclusive assignments. For those molecules, we assigned all carbon-carbon bonds as single bonds.
MLWF centers nearest H in C-H bonds were associated to that bond, and MLWF centers nearest the O in C-O-C, C=O, C$\equiv$O (carbon monoxide), C-O$^-$ and C-OH were associated to that functional group.
Unassigned centers nearest to carbon were considered lone pairs,
although these were found only transiently in simulations.

The total electronic charge in C-O and C-H bonds,
as determined from the MLWF centers,
was divided between the carbon atom and the bonded partner in the usual way.
For example, in C=O bonds, there are 4 electrons in the bond and two lone pairs on the oxygen atom.
Since O has an oxidation state of -2, all electronic charge in the bond is assigned to the oxygen, and none to the carbon atom.
Electrons in carbon-carbon bonds were assumed to be shared equally between C atoms.
Lone pairs on carbon atoms contributed -2$e$ of negative charge to the local charge of the carbon.
Summing these contributions yielded $q_{\rm local}$ for each carbon atom, and the oxidation state was computed as
$\left(q_{\rm local} + 4e\right)/e$.

\section*{Supporting information}
Details of simulations, energy in simulations, covalent bond length cutoff distances, change in the number of bonds per carbon atom, illustrated examples of reaction mechanisms.

\section*{Acknowledgements}
This work was supported by the Hong Kong Research Grants Council (RGC) (Projects GRF-16301723, GRF-16306621, GRF-16302423, and GRF-16310225), and National Natural Science Foundation of China/RGC Joint Research Scheme (N\_HKUST664/24).
Part of this work was carried out using computational resources from the National Supercomputer Center in Guangzhou, China. 

\bibliography{ref}

\begin{thebibliography}{10}
\expandafter\ifx\csname url\endcsname\relax
  \def\url#1{\texttt{#1}}\fi
\expandafter\ifx\csname urlprefix\endcsname\relax\def\urlprefix{URL }\fi
\providecommand{\bibinfo}[2]{#2}
\providecommand{\eprint}[2][]{\url{#2}}

\bibitem{Sephton2013On}
\bibinfo{author}{Sephton, M.~A.} \& \bibinfo{author}{Hazen, R.~M.}
\newblock \bibinfo{title}{On the origins of deep hydrocarbons}.
\newblock \emph{\bibinfo{journal}{Rev. Mineral. Geochem.}} \textbf{\bibinfo{volume}{75}}, \bibinfo{pages}{449--465} (\bibinfo{year}{2013}).

\bibitem{McCollom2013Laboratory}
\bibinfo{author}{McCollom, T.~M.}
\newblock \bibinfo{title}{{Laboratory simulations of abiotic hydrocarbon formation in Earth's deep subsurface}}.
\newblock \emph{\bibinfo{journal}{Rev. Mineral. Geochem.}} \textbf{\bibinfo{volume}{75}}, \bibinfo{pages}{467--494} (\bibinfo{year}{2013}).

\bibitem{vitale2017massive}
\bibinfo{author}{Vitale~Brovarone, A.} \emph{et~al.}
\newblock \bibinfo{title}{Massive production of abiotic methane during subduction evidenced in metamorphosed ophicarbonates from the {Italian Alps}}.
\newblock \emph{\bibinfo{journal}{Nat. Commun.}} \textbf{\bibinfo{volume}{8}}, \bibinfo{pages}{14134} (\bibinfo{year}{2017}).

\bibitem{Tao2018Formation}
\bibinfo{author}{Tao, R.} \emph{et~al.}
\newblock \bibinfo{title}{{Formation of abiotic hydrocarbon from reduction of carbonate in subduction zones: Constraints from petrological observation and experimental simulation}}.
\newblock \emph{\bibinfo{journal}{Geochim. Cosmochim. Acta}} \textbf{\bibinfo{volume}{239}}, \bibinfo{pages}{390--408} (\bibinfo{year}{2018}).

\bibitem{Sverjensky2020Changing}
\bibinfo{author}{Sverjensky, D.~A.}, \bibinfo{author}{Daniel, I.} \& \bibinfo{author}{Brovarone, A.~V.}
\newblock \bibinfo{title}{The changing character of carbon in fluids with pressure: {O}rganic geochemistry of {E}arth's upper mantle fluids}.
\newblock In \bibinfo{editor}{Manning, C.~E.}, \bibinfo{editor}{Lin, J.-F.} \& \bibinfo{editor}{Mao, W.~L.} (eds.) \emph{\bibinfo{booktitle}{Carbon in Earth's Interior}}, chap.~\bibinfo{chapter}{22}, \bibinfo{pages}{259--269} (\bibinfo{publisher}{John Wiley {\&} Sons, Inc}, \bibinfo{year}{2020}).

\bibitem{vitale2020subduction}
\bibinfo{author}{Vitale~Brovarone, A.} \emph{et~al.}
\newblock \bibinfo{title}{Subduction hides high-pressure sources of energy that may feed the deep subsurface biosphere}.
\newblock \emph{\bibinfo{journal}{Nat. Commun.}} \textbf{\bibinfo{volume}{11}}, \bibinfo{pages}{3880} (\bibinfo{year}{2020}).

\bibitem{pena2021situ}
\bibinfo{author}{Pe{\~n}a-Alvarez, M.} \emph{et~al.}
\newblock \bibinfo{title}{In-situ abiogenic methane synthesis from diamond and graphite under geologically relevant conditions}.
\newblock \emph{\bibinfo{journal}{Nat. Commun.}} \textbf{\bibinfo{volume}{12}}, \bibinfo{pages}{6387} (\bibinfo{year}{2021}).

\bibitem{Kelemen2015Reevaluating}
\bibinfo{author}{Kelemen, P.~B.} \& \bibinfo{author}{Manning, C.~E.}
\newblock \bibinfo{title}{Reevaluating carbon fluxes in subduction zones, what goes down, mostly comes up}.
\newblock \emph{\bibinfo{journal}{Proc. Natl. Acad. Sci. U. S. A.}} \textbf{\bibinfo{volume}{112}}, \bibinfo{pages}{E3997--E4006} (\bibinfo{year}{2015}).

\bibitem{Kutcherov2020Hydrocarbon}
\bibinfo{author}{Kutcherov, V.~G.}, \bibinfo{author}{Ivanov, K.}, \bibinfo{author}{Mukhina, E.} \& \bibinfo{author}{Serovaiskii, A.}
\newblock \bibinfo{title}{{Deep hydrocarbon cycle: An experimental simulation}}.
\newblock In \bibinfo{editor}{Manning, C.~E.}, \bibinfo{editor}{Lin, J.-F.} \& \bibinfo{editor}{Mao, W.~L.} (eds.) \emph{\bibinfo{booktitle}{Carbon in Earth's Interior}}, chap.~\bibinfo{chapter}{26}, \bibinfo{pages}{329--339} (\bibinfo{publisher}{John Wiley {\&} Sons, Inc.}, \bibinfo{year}{2020}), \bibinfo{edition}{1} edn.

\bibitem{Hazen2013Deep}
\bibinfo{author}{Hazen, R.~M.} \& \bibinfo{author}{Schiffries, C.~M.}
\newblock \bibinfo{title}{Why deep carbon?}
\newblock \emph{\bibinfo{journal}{Rev. Mineral. Geochem.}} \textbf{\bibinfo{volume}{75}}, \bibinfo{pages}{1--6} (\bibinfo{year}{2013}).

\bibitem{Mendeleev1877}
\bibinfo{author}{Mendeleev, D.}
\newblock \bibinfo{title}{L'origine du p\'{e}trole}.
\newblock \emph{\bibinfo{journal}{Rev. Sci.}} \textbf{\bibinfo{volume}{8}}, \bibinfo{pages}{409–416} (\bibinfo{year}{1877}).

\bibitem{Gold1980Deep}
\bibinfo{author}{Gold, T.} \& \bibinfo{author}{Soter, S.}
\newblock \bibinfo{title}{{The deep-Earth-gas hypothesis}}.
\newblock \emph{\bibinfo{journal}{Sci. Am.}} \textbf{\bibinfo{volume}{242}}, \bibinfo{pages}{154--161} (\bibinfo{year}{1980}).

\bibitem{Kolesnikov2009Methane}
\bibinfo{author}{Kolesnikov, A.}, \bibinfo{author}{Kutcherov, V.~G.} \& \bibinfo{author}{Goncharov, A.~F.}
\newblock \bibinfo{title}{{Methane-derived hydrocarbons produced under upper-mantle conditions}}.
\newblock \emph{\bibinfo{journal}{Nat. Geosci.}} \textbf{\bibinfo{volume}{2}}, \bibinfo{pages}{566--570} (\bibinfo{year}{2009}).

\bibitem{kutcherov2010synthesis}
\bibinfo{author}{Kutcherov, V.}, \bibinfo{author}{Kolesnikov, A.}, \bibinfo{author}{Dyuzheva, T.} \& \bibinfo{author}{Brazhkin, V.}
\newblock \bibinfo{title}{Synthesis of hydrocarbons under upper mantle conditions: Evidence for the theory of abiotic deep petroleum origin}.
\newblock In \emph{\bibinfo{booktitle}{J. Phys.: Conf. Ser.}}, vol. \bibinfo{volume}{215}, \bibinfo{pages}{012103} (\bibinfo{organization}{IOP Publishing}, \bibinfo{year}{2010}).

\bibitem{kolesnikov2017chemistry}
\bibinfo{author}{Kolesnikov, A.~Y.}, \bibinfo{author}{Saul, J.~M.} \& \bibinfo{author}{Kutcherov, V.~G.}
\newblock \bibinfo{title}{Chemistry of hydrocarbons under extreme thermobaric conditions}.
\newblock \emph{\bibinfo{journal}{ChemistrySelect}} \textbf{\bibinfo{volume}{2}}, \bibinfo{pages}{1336--1352} (\bibinfo{year}{2017}).

\bibitem{Brady1981Mechanism}
\bibinfo{author}{{Brady III}, R.~C.} \& \bibinfo{author}{Pettit, R.}
\newblock \bibinfo{title}{{On the mechanism of the Fischer-Tropsch reaction. The chain propagation step}}.
\newblock \emph{\bibinfo{journal}{J. Am. Chem. Soc.}} \textbf{\bibinfo{volume}{103}}, \bibinfo{pages}{1287--1289} (\bibinfo{year}{1981}).

\bibitem{Jiang2024Role}
\bibinfo{author}{Jiang, L.} \emph{et~al.}
\newblock \bibinfo{title}{{Role of H$_2$O in Catalytic Conversion of C$_1$ Molecules}}.
\newblock \emph{\bibinfo{journal}{J. Am. Chem. Soc.}} \textbf{\bibinfo{volume}{146}}, \bibinfo{pages}{2857--2875} (\bibinfo{year}{2024}).

\bibitem{Thompson1992Water}
\bibinfo{author}{Thompson, A.~B.}
\newblock \bibinfo{title}{{Water in the Earth's upper mantle}}.
\newblock \emph{\bibinfo{journal}{Nature}} \textbf{\bibinfo{volume}{358}}, \bibinfo{pages}{295--302} (\bibinfo{year}{1992}).

\bibitem{Manning2013Chemistry}
\bibinfo{author}{Manning, C.~E.}, \bibinfo{author}{Shock, E.~L.} \& \bibinfo{author}{Sverjensky, D.~A.}
\newblock \bibinfo{title}{{The chemistry of carbon in aqueous fluids at crustal and upper-mantle conditions: Experimental and theoretical constraints}}.
\newblock \emph{\bibinfo{journal}{Rev. Mineral. Geochem.}} \textbf{\bibinfo{volume}{75}}, \bibinfo{pages}{109--148} (\bibinfo{year}{2013}).

\bibitem{Tschauner2018Ice-VII}
\bibinfo{author}{Tschauner, O.} \emph{et~al.}
\newblock \bibinfo{title}{{Ice-VII inclusions in diamonds: Evidence for aqueous fluid in Earth's deep mantle}}.
\newblock \emph{\bibinfo{journal}{Science}} \textbf{\bibinfo{volume}{359}}, \bibinfo{pages}{1136--1139} (\bibinfo{year}{2018}).

\bibitem{bali2013water}
\bibinfo{author}{Bali, E.}, \bibinfo{author}{Aud{\'e}tat, A.} \& \bibinfo{author}{Keppler, H.}
\newblock \bibinfo{title}{Water and hydrogen are immiscible in earth’s mantle}.
\newblock \emph{\bibinfo{journal}{Nature}} \textbf{\bibinfo{volume}{495}}, \bibinfo{pages}{220--222} (\bibinfo{year}{2013}).

\bibitem{Zhang2009Model}
\bibinfo{author}{Zhang, C.} \& \bibinfo{author}{Duan, Z.}
\newblock \bibinfo{title}{A model for {C--O--H} fluid in the {E}arth's mantle}.
\newblock \emph{\bibinfo{journal}{Geochim. Cosmochim. Acta}} \textbf{\bibinfo{volume}{73}}, \bibinfo{pages}{2089--2102} (\bibinfo{year}{2009}).

\bibitem{Pan2013Dielectric}
\bibinfo{author}{Pan, D.}, \bibinfo{author}{Spanu, L.}, \bibinfo{author}{Harrison, B.}, \bibinfo{author}{Sverjensky, D.~A.} \& \bibinfo{author}{Galli, G.}
\newblock \bibinfo{title}{Dielectric properties of water under extreme conditions and transport of carbonates in the deep {Earth}}.
\newblock \emph{\bibinfo{journal}{Proc. Natl. Acad. Sci. U. S. A.}} \textbf{\bibinfo{volume}{110}}, \bibinfo{pages}{6646--6650} (\bibinfo{year}{2013}).

\bibitem{Sverjensky2014Important}
\bibinfo{author}{Sverjensky, D.~A.}, \bibinfo{author}{Stagno, V.} \& \bibinfo{author}{Huang, F.}
\newblock \bibinfo{title}{{Important role for organic carbon in subduction-zone fluids in the deep carbon cycle}}.
\newblock \emph{\bibinfo{journal}{Nat. Geosci.}} \textbf{\bibinfo{volume}{7}}, \bibinfo{pages}{909--913} (\bibinfo{year}{2014}).

\bibitem{Pan2016Fate}
\bibinfo{author}{Pan, D.} \& \bibinfo{author}{Galli, G.}
\newblock \bibinfo{title}{The fate of carbon dioxide in water-rich fluids under extreme conditions}.
\newblock \emph{\bibinfo{journal}{Sci. Adv.}} \textbf{\bibinfo{volume}{2}}, \bibinfo{pages}{e1601278} (\bibinfo{year}{2016}).

\bibitem{Abramson2017Water-carbon}
\bibinfo{author}{Abramson, E.~H.}, \bibinfo{author}{Bollengier, O.} \& \bibinfo{author}{Brown, J.~M.}
\newblock \bibinfo{title}{{The water-carbon dioxide miscibility surface to 450 {$^{\circ}$}C and 7 GPa}}.
\newblock \emph{\bibinfo{journal}{Am. J. Sci.}} \textbf{\bibinfo{volume}{317}}, \bibinfo{pages}{967--989} (\bibinfo{year}{2017}).

\bibitem{Stolte2019Large}
\bibinfo{author}{Stolte, N.} \& \bibinfo{author}{Pan, D.}
\newblock \bibinfo{title}{{Large presence of carbonic acid in CO$_2$-rich aqueous fluids under Earth's mantle conditions}}.
\newblock \emph{\bibinfo{journal}{J. Phys. Chem. Lett.}} \textbf{\bibinfo{volume}{10}}, \bibinfo{pages}{5135--5141} (\bibinfo{year}{2019}).

\bibitem{dettori2020carbon}
\bibinfo{author}{Dettori, R.} \& \bibinfo{author}{Donadio, D.}
\newblock \bibinfo{title}{Carbon dioxide, bicarbonate and carbonate ions in aqueous solutions under deep {E}arth conditions}.
\newblock \emph{\bibinfo{journal}{Phys. Chem. Chem. Phys.}} \textbf{\bibinfo{volume}{22}}, \bibinfo{pages}{10717--10725} (\bibinfo{year}{2020}).

\bibitem{stolte2021water}
\bibinfo{author}{Stolte, N.}, \bibinfo{author}{Yu, J.}, \bibinfo{author}{Chen, Z.}, \bibinfo{author}{Sverjensky, D.~A.} \& \bibinfo{author}{Pan, D.}
\newblock \bibinfo{title}{Water--gas shift reaction produces formate at extreme pressures and temperatures in deep earth fluids}.
\newblock \emph{\bibinfo{journal}{J. Phys. Chem. Lett.}} \textbf{\bibinfo{volume}{12}}, \bibinfo{pages}{4292--4298} (\bibinfo{year}{2021}).

\bibitem{stolte2022nanoconfinement}
\bibinfo{author}{Stolte, N.}, \bibinfo{author}{Hou, R.} \& \bibinfo{author}{Pan, D.}
\newblock \bibinfo{title}{Nanoconfinement facilitates reactions of carbon dioxide in supercritical water}.
\newblock \emph{\bibinfo{journal}{Nat. Commun.}} \textbf{\bibinfo{volume}{13}}, \bibinfo{pages}{5932} (\bibinfo{year}{2022}).

\bibitem{li2024synthesis}
\bibinfo{author}{Li, T.} \emph{et~al.}
\newblock \bibinfo{title}{Synthesis and stability of biomolecules in {C--H--O--N} fluids under earth’s upper mantle conditions}.
\newblock \emph{\bibinfo{journal}{J. Am. Chem. Soc.}} \textbf{\bibinfo{volume}{146}}, \bibinfo{pages}{31240--31250} (\bibinfo{year}{2024}).

\bibitem{fei2025abiotic}
\bibinfo{author}{Fei, C.}, \bibinfo{author}{Guo, S.}, \bibinfo{author}{Li, Y.} \& \bibinfo{author}{Liu, J.}
\newblock \bibinfo{title}{Abiotic synthesis during the interaction of ferrous chloride--rich silicic fluids with marble under high-grade metamorphic conditions}.
\newblock \emph{\bibinfo{journal}{Proc. Natl. Acad. Sci. U. S. A.}} \textbf{\bibinfo{volume}{122}}, \bibinfo{pages}{e2423043122} (\bibinfo{year}{2025}).

\bibitem{Lobanov2013Carbon}
\bibinfo{author}{Lobanov, S.~S.} \emph{et~al.}
\newblock \bibinfo{title}{Carbon precipitation from heavy hydrocarbon fluid in deep planetary interiors}.
\newblock \emph{\bibinfo{journal}{Nat. Commun.}} \textbf{\bibinfo{volume}{4}}, \bibinfo{pages}{2446} (\bibinfo{year}{2013}).

\bibitem{Gygi2005Abinitio}
\bibinfo{author}{Gygi, F.} \& \bibinfo{author}{Galli, G.}
\newblock \bibinfo{title}{{Ab initio simulation in extreme conditions}}.
\newblock \emph{\bibinfo{journal}{Mater. Today}} \textbf{\bibinfo{volume}{8}}, \bibinfo{pages}{26--32} (\bibinfo{year}{2005}).

\bibitem{Spanu2011Stability}
\bibinfo{author}{Spanu, L.}, \bibinfo{author}{Donadio, D.}, \bibinfo{author}{Hohl, D.}, \bibinfo{author}{Schwegler, E.} \& \bibinfo{author}{Galli, G.}
\newblock \bibinfo{title}{{Stability of hydrocarbons at deep Earth pressures and temperatures}}.
\newblock \emph{\bibinfo{journal}{Proc. Natl. Acad. Sci. U. S. A.}} \textbf{\bibinfo{volume}{108}}, \bibinfo{pages}{6843--6846} (\bibinfo{year}{2011}).

\bibitem{kuang2022high}
\bibinfo{author}{Kuang, H.} \& \bibinfo{author}{Tse, J.~S.}
\newblock \bibinfo{title}{{High-Temperature}, {High-Pressure} reactions of {H$_2$} with {CaCO$_3$} melts}.
\newblock \emph{\bibinfo{journal}{Phys. Status Solidi B}} \textbf{\bibinfo{volume}{259}}, \bibinfo{pages}{2100644} (\bibinfo{year}{2022}).

\bibitem{ancilotto1997dissociation}
\bibinfo{author}{Ancilotto, F.}, \bibinfo{author}{Chiarotti, G.~L.}, \bibinfo{author}{Scandolo, S.} \& \bibinfo{author}{Tosatti, E.}
\newblock \bibinfo{title}{Dissociation of methane into hydrocarbons at extreme (planetary) pressure and temperature}.
\newblock \emph{\bibinfo{journal}{Science}} \textbf{\bibinfo{volume}{275}}, \bibinfo{pages}{1288--1290} (\bibinfo{year}{1997}).

\bibitem{lee2011mixtures}
\bibinfo{author}{Lee, M.-S.} \& \bibinfo{author}{Scandolo, S.}
\newblock \bibinfo{title}{Mixtures of planetary ices at extreme conditions}.
\newblock \emph{\bibinfo{journal}{Nat. Commun.}} \textbf{\bibinfo{volume}{2}}, \bibinfo{pages}{185} (\bibinfo{year}{2011}).

\bibitem{li2011quantum}
\bibinfo{author}{Li, D.}, \bibinfo{author}{Zhang, P.} \& \bibinfo{author}{Yan, J.}
\newblock \bibinfo{title}{Quantum molecular dynamics simulations for the nonmetal-metal transition in shocked methane}.
\newblock \emph{\bibinfo{journal}{Phys. Rev. B}} \textbf{\bibinfo{volume}{84}}, \bibinfo{pages}{184204} (\bibinfo{year}{2011}).

\bibitem{Schroder1998Ethylenedione}
\bibinfo{author}{Schr{\"{o}}der, D.} \emph{et~al.}
\newblock \bibinfo{title}{{Ethylenedione: An intrinsically short-lived molecule}}.
\newblock \emph{\bibinfo{journal}{Chem. Eur. J.}} \textbf{\bibinfo{volume}{4}}, \bibinfo{pages}{2550--2557} (\bibinfo{year}{1998}).

\bibitem{Xin2019Global}
\bibinfo{author}{Xin, J.-F.}, \bibinfo{author}{Han, X.-R.}, \bibinfo{author}{He, F.-F.} \& \bibinfo{author}{Ding, Y.-H.}
\newblock \bibinfo{title}{{Global isomeric survey of elusive cyclopropanetrione: Unknown but viable isomers}}.
\newblock \emph{\bibinfo{journal}{Front. Chem.}} \textbf{\bibinfo{volume}{7}} (\bibinfo{year}{2019}).

\bibitem{Mato2020Stability}
\bibinfo{author}{Mato, J.}, \bibinfo{author}{Poole, D.} \& \bibinfo{author}{Gordon, M.~S.}
\newblock \bibinfo{title}{{Stability and dissociation of ethylenedione (OCCO)}}.
\newblock \emph{\bibinfo{journal}{J. Phys. Chem. A}} \textbf{\bibinfo{volume}{124}}, \bibinfo{pages}{8209--8222} (\bibinfo{year}{2020}).

\bibitem{Gleiter1995Stability}
\bibinfo{author}{Gleiter, R.}, \bibinfo{author}{Hyla-Kryspin, I.} \& \bibinfo{author}{Pfeifer, K.~H.}
\newblock \bibinfo{title}{{On the stability of the tetramers of carbon monoxide, hydrogen isocyanide, and vinylidene. A molecular orbital theoretical rationalization}}.
\newblock \emph{\bibinfo{journal}{J. Org. Chem.}} \textbf{\bibinfo{volume}{60}}, \bibinfo{pages}{5878--5883} (\bibinfo{year}{1995}).

\bibitem{Guo2012Probing}
\bibinfo{author}{Guo, J.~C.}, \bibinfo{author}{Hou, G.~L.}, \bibinfo{author}{Li, S.~D.} \& \bibinfo{author}{Wang, X.~B.}
\newblock \bibinfo{title}{{Probing the low-lying electronic states of cyclobutanetetraone (C$_4$O$_4$) and its radical anion: A low-temperature anion photoelectron spectroscopic approach}}.
\newblock \emph{\bibinfo{journal}{J. Phys. Chem. Lett.}} \textbf{\bibinfo{volume}{3}}, \bibinfo{pages}{304--308} (\bibinfo{year}{2012}).

\bibitem{mccollom1999lipid}
\bibinfo{author}{McCollom, T.~M.}, \bibinfo{author}{Ritter, G.} \& \bibinfo{author}{Simoneit, B.~R.}
\newblock \bibinfo{title}{{Lipid synthesis under hydrothermal conditions by Fischer-Tropsch-type reactions}}.
\newblock \emph{\bibinfo{journal}{Origins Life Evol. B.}} \textbf{\bibinfo{volume}{29}}, \bibinfo{pages}{153--166} (\bibinfo{year}{1999}).

\bibitem{McCollom2006Carbon}
\bibinfo{author}{McCollom, T.~M.} \& \bibinfo{author}{Seewald, J.~S.}
\newblock \bibinfo{title}{{Carbon isotope composition of organic compounds produced by abiotic synthesis under hydrothermal conditions}}.
\newblock \emph{\bibinfo{journal}{Earth Planet. Sci. Lett.}} \textbf{\bibinfo{volume}{243}}, \bibinfo{pages}{74--84} (\bibinfo{year}{2006}).

\bibitem{Plank2019Subducting}
\bibinfo{author}{Plank, T.} \& \bibinfo{author}{Manning, C.~E.}
\newblock \bibinfo{title}{Subducting carbon}.
\newblock \emph{\bibinfo{journal}{Nature}} \textbf{\bibinfo{volume}{574}}, \bibinfo{pages}{343--352} (\bibinfo{year}{2019}).

\bibitem{giuntoli2024methane}
\bibinfo{author}{Giuntoli, F.} \emph{et~al.}
\newblock \bibinfo{title}{Methane-hydrogen-rich fluid migration may trigger seismic failure in subduction zones at forearc depths}.
\newblock \emph{\bibinfo{journal}{Nat. Commun.}} \textbf{\bibinfo{volume}{15}}, \bibinfo{pages}{480} (\bibinfo{year}{2024}).

\bibitem{dobe2025tracking}
\bibinfo{author}{Dobe, R.} \emph{et~al.}
\newblock \bibinfo{title}{Tracking molecular hydrogen migration along a subduction shear zone}.
\newblock \emph{\bibinfo{journal}{Geol. Soc. Am. Bull.}} \textbf{\bibinfo{volume}{137}}, \bibinfo{pages}{5241--5264} (\bibinfo{year}{2025}).

\bibitem{Gygi2008Architecture}
\bibinfo{author}{Gygi, F.}
\newblock \bibinfo{title}{{Architecture of Qbox: A scalable first-principles molecular dynamics code}}.
\newblock \emph{\bibinfo{journal}{IBM J. Res. Dev.}} \textbf{\bibinfo{volume}{52}}, \bibinfo{pages}{137--144} (\bibinfo{year}{2008}).

\bibitem{Perdew1996Generalized}
\bibinfo{author}{Perdew, J.~P.}, \bibinfo{author}{Burke, K.} \& \bibinfo{author}{Ernzerhof, M.}
\newblock \bibinfo{title}{{Generalized gradient approximation made simple}}.
\newblock \emph{\bibinfo{journal}{Phys. Rev. Lett.}} \textbf{\bibinfo{volume}{77}}, \bibinfo{pages}{3865--3868} (\bibinfo{year}{1996}).

\bibitem{Adamo1999Toward}
\bibinfo{author}{Adamo, C.} \& \bibinfo{author}{Barone, V.}
\newblock \bibinfo{title}{{Toward reliable density functional methods without adjustable parameters: The PBE0 model}}.
\newblock \emph{\bibinfo{journal}{J. Chem. Phys.}} \textbf{\bibinfo{volume}{110}}, \bibinfo{pages}{6158--6170} (\bibinfo{year}{1999}).

\bibitem{Pan2014Refractive}
\bibinfo{author}{Pan, D.}, \bibinfo{author}{Wan, Q.} \& \bibinfo{author}{Galli, G.}
\newblock \bibinfo{title}{The refractive index and electronic gap of water and ice increase with increasing pressure}.
\newblock \emph{\bibinfo{journal}{Nat. Commun.}} \textbf{\bibinfo{volume}{5}}, \bibinfo{pages}{3919} (\bibinfo{year}{2014}).

\bibitem{grimme2010consistent}
\bibinfo{author}{Grimme, S.}, \bibinfo{author}{Antony, J.}, \bibinfo{author}{Ehrlich, S.} \& \bibinfo{author}{Krieg, H.}
\newblock \bibinfo{title}{{A consistent and accurate ab initio parametrization of density functional dispersion correction (DFT-D) for the 94 elements H-Pu}}.
\newblock \emph{\bibinfo{journal}{J. Chem. Phys.}} \textbf{\bibinfo{volume}{132}}, \bibinfo{pages}{154104} (\bibinfo{year}{2010}).

\bibitem{hammer1999improved}
\bibinfo{author}{Hammer, B.}, \bibinfo{author}{Hansen, L.~B.} \& \bibinfo{author}{N{\o}rskov, J.~K.}
\newblock \bibinfo{title}{{Improved adsorption energetics within density-functional theory using revised Perdew-Burke-Ernzerhof functionals}}.
\newblock \emph{\bibinfo{journal}{Phys. Rev. B}} \textbf{\bibinfo{volume}{59}}, \bibinfo{pages}{7413--7421} (\bibinfo{year}{1999}).

\bibitem{Hamann2013Optimized}
\bibinfo{author}{Hamann, D.~R.}
\newblock \bibinfo{title}{{Optimized norm-conserving Vanderbilt pseudopotentials}}.
\newblock \emph{\bibinfo{journal}{Phys. Rev. B}} \textbf{\bibinfo{volume}{88}}, \bibinfo{pages}{085117} (\bibinfo{year}{2013}).

\bibitem{Schlipf2015Optimization}
\bibinfo{author}{Schlipf, M.} \& \bibinfo{author}{Gygi, F.}
\newblock \bibinfo{title}{{Optimization algorithm for the generation of ONCV pseudopotentials}}.
\newblock \emph{\bibinfo{journal}{Comput. Phys. Commun.}} \textbf{\bibinfo{volume}{196}}, \bibinfo{pages}{36--44} (\bibinfo{year}{2015}).

\bibitem{Bussi2007Canonical}
\bibinfo{author}{Bussi, G.}, \bibinfo{author}{Donadio, D.} \& \bibinfo{author}{Parrinello, M.}
\newblock \bibinfo{title}{{Canonical sampling through velocity rescaling}}.
\newblock \emph{\bibinfo{journal}{J. Chem. Phys.}} \textbf{\bibinfo{volume}{126}}, \bibinfo{pages}{014101} (\bibinfo{year}{2007}).

\bibitem{focher1994structural}
\bibinfo{author}{Focher, P.}, \bibinfo{author}{Chiarotti, G.~L.}, \bibinfo{author}{Bernasconi, M.}, \bibinfo{author}{Tosatti, E.} \& \bibinfo{author}{Parrinello, M.}
\newblock \bibinfo{title}{Structural phase transformations via first-principles simulation}.
\newblock \emph{\bibinfo{journal}{Europhys. Lett.}} \textbf{\bibinfo{volume}{26}}, \bibinfo{pages}{345} (\bibinfo{year}{1994}).

\bibitem{Marzari2012Maximally}
\bibinfo{author}{Marzari, N.}, \bibinfo{author}{Mostofi, A.~A.}, \bibinfo{author}{Yates, J.~R.}, \bibinfo{author}{Souza, I.} \& \bibinfo{author}{Vanderbilt, D.}
\newblock \bibinfo{title}{{Maximally localized Wannier functions: Theory and applications}}.
\newblock \emph{\bibinfo{journal}{Rev. Mod. Phys.}} \textbf{\bibinfo{volume}{84}}, \bibinfo{pages}{1419--1475} (\bibinfo{year}{2012}).

\end{thebibliography}
\clearpage

\end{document}